\documentclass[12pt,journal,onecolumn]{IEEEtran}
\usepackage[hang]{subfigure}
\usepackage{epsfig,rotating,setspace,latexsym,amsmath,epsf,amssymb,amsfonts,bm,theorem,epstopdf}
\usepackage{cite,authblk}
\usepackage{algorithm}
\usepackage[noend]{algpseudocode}
\usepackage{setspace}
\usepackage{color}
\usepackage{graphicx}
\usepackage{mathrsfs}
\ifCLASSOPTIONcompsoc
\usepackage[caption=false, font=normalsize, labelfont=sf, textfont=sf]{subfig}
\else
\usepackage[caption=false, font=footnotesize]{subfig}
\fi

\newtheorem{lemma}{Lemma}

\newenvironment{Proof}[1]{\medskip\par\noindent{\bf Proof:\,}\,#1}{{\mbox{\,$\blacksquare$}\par}}

\allowdisplaybreaks

\definecolor{green1}{rgb}{0.2,0.7,0.2}
\definecolor{brown}{rgb}{1,0.5,0.2}

\usepackage[colorinlistoftodos,textwidth=\marginparwidth]{todonotes}

\begin{document}

\title{The Role of Gossiping for Information Dissemination over Networked Agents}
\author{ Melih Bastopcu, S. Rasoul Etesami, and Tamer Ba\c{s}ar\\
	\normalsize Coordinated Science Laboratory\\
	\normalsize University of Illinois Urbana-Champaign, Urbana, IL, 61801\\
	\normalsize  \texttt{{\small (bastopcu,etesami1,basar1)@illinois.edu}}
	\thanks{*This work was supported in part by the NSF CAREER Award under Grant EPCN-1944403, and in part by the ARO MURI Grant AG285.}}
\maketitle
\vspace{-1.5cm}
\begin{abstract}
We consider information dissemination over a network of gossiping agents (nodes). In this model, a source keeps the most up-to-date information about a time-varying binary state of the world, and $n$ receiver nodes want to follow the information at the source as accurately as possible. When the information at the source changes, the source first sends updates to a subset of $m\leq n$ nodes. After that, the nodes share their local information during the \textit{gossiping period} to disseminate the information further. The nodes then estimate the information at the source using the majority rule at the end of the gossiping period. To analyze information dissemination, we introduce a new error metric to find the average percentage of nodes that can accurately obtain the most up-to-date information at the source. We characterize the equations necessary to obtain the steady-state distribution for the average error and then analyze the system behavior under both high and low gossip rates. In the high gossip rate, in which each node can access other nodes' information more frequently, we show that all the nodes start to behave like a single node and update their information based on the majority of the information in the network. In the low gossip rate, we introduce and analyze the so-called \textit{gossip gain}, which is the reduction at the average error due to gossiping. In particular, we develop an adaptive policy that the source can use to determine its current transmission capacity $m$ based on its past transmission rates and the accuracy of the information at the nodes. Through numerical results, we first show that when the source's transmission capacity $m$ is limited, gossiping can be harmful as it causes incorrect information to disseminate. We then find the optimal gossip rates to minimize the average error for a fixed $m$. Finally, we illustrate the outperformance of our adaptive policy compared to the constant $m$-selection policy even for the high gossip rates.             
\end{abstract}
 
\section{Introduction}
Motivated by many applications such as autonomous vehicular systems, content advertising on social media, and city emergency warning systems, information dissemination over the networks has gained significant attention. For instance, in the case of autonomous vehicular systems or city emergency warning systems, timely-critical information such as accident alerts or tornado warnings need to be disseminated as quickly and accurately as possible. As another example, companies often want to let their potential customers know about their latest products through advertisements over social media. In both of these examples, there is a single information source where the most up-to-date information is disseminated to multiple receivers over time.

In this paper, we consider a communication system with a source and $n$ receiver nodes. The source keeps the most recent information about the state of the world, which takes values $0$ or $1$, and changes according to an exponential distribution. Upon each information update, the source wants to let the receiver nodes know about the most recent information. As the source has limited transmission capacity, it cannot send information to more than $m\leq n$ nodes, and each information transmission at the source takes an exponentially distributed length of time. After sending updates to $m$ nodes, in order to further disseminate information, each pair of receiver nodes share their local information between each other, a process we shall refer to as \textit{gossiping}. The gossiping period continues until the information at the source is updated again. At the end of each gossiping period, each receiver node that did not get the most recent information directly from the source comes up with an estimate based on the majority of the information it received from the other nodes. In order to measure the accuracy of the information dissemination at the end of each update cycle, we consider an error metric that takes value 1 for a receiver node that has a different estimate compared to the information at the source. 

\subsection{Related Work}
In gossip network literature, the model where only one node tries to spread its information to the entire network has been considered in \cite{shah2009gossip} and named \textit{single-piece
dissemination}. The multi-piece spreading where all nodes try to spread their individual information to the remaining nodes has been studied in \cite{mosk2006}. Moreover, the problem of finding the average of all nodes' initial information on a gossip network has been studied under the framework of \textit{distributed averaging} in \cite{Boydgossip2006a, Boydgossip2006b}. The main goal of these works has been to analytically characterize either the information spreading time \cite{shah2009gossip, mosk2006} or the averaging time \cite{Boydgossip2006a, Boydgossip2006b} in the entire network. In another line of research, to measure the timeliness of information, age of information has been proposed in \cite{Kaul12a} and it has been extensively studied in multi-hop multicast networks \cite{Talak17, Tripathi17, Bedewy17b, Zhong17a, Zhong18b, Buyukates18b, Krishnan19, Farazi19}, content freshness in the web \cite{Cho03, Azar18, Kolobov19a, Brewington00}, and timely remote estimation of random processes \cite{Wang19a, Sun17b, Sun18b, Chakravorty18, Kam20a, arafa2020, Bastopcu20a}. For a more detailed review of the age of information, we refer to \cite{Kosta17c, Yates20a}. Recently, scaling of the age of information has been considered in gossip networks \cite{yates2021age, Bastopcu21c, Buyukates21a, buyukates21e}.

Different from the earlier works on gossip networks as in \cite{shah2009gossip, mosk2006}, we consider here a time-varying information source. Moreover, instead of tracking the information spreading time, we study the average percentage of the nodes that have access to the most recent information at the source before it is updated. Compared to \cite{shah2009gossip, mosk2006}, our information updating is different and consists of two phases, where in the first phase, only the source can send updates to $m$ nodes, and in the second phase, i.e., in the gossiping phase, only the nodes can share their local information. Thus, in the gossiping phase, incorrect information in the network can also spread. The works \cite{yates2021age, Bastopcu21c, Buyukates21a, buyukates21e} have considered the age of information in gossip networks where each information update at the source is treated as a new update and content of the information has not been considered. In this work, we consider a binary information source that changes its state based on Poisson updates. Furthermore, in \cite{yates2021age, Bastopcu21c, Buyukates21a, buyukates21e}, the nodes update their information only if they receive fresher information. In contrast, in our work, the nodes that do not receive any update directly from the source make decisions based on the majority of the updates that they receive from the other nodes. As a result, the error metric and the information updating model that we consider differs from the earlier works in \cite{yates2021age, Bastopcu21c, Buyukates21a, buyukates21e}. 

\subsection{Contributions}

In this work, we first characterize the equations necessary to obtain the steady-state distribution of the average error. Then, we provide analytical results for the high and low gossip rates. When the gossip rate is high, we show that the probability of obtaining correct information converges to a step function where if the majority of the nodes have the correct information, all the nodes are able to estimate the information correctly with probability 1. In other words, as the gossip rate increases, information at all nodes becomes available to each other, and all the nodes in the network behave like a single node. However, when the gossip rate is low, the gossiping phase can be approximated by either not receiving any updates, in which case the nodes hold on to their prior information, or receiving a single update. Based on this approximation, we characterize analytically the gain obtained through gossiping and find an adaptive selection policy for the source, which suggests that the source should send updates to more nodes when the nodes have mostly incorrect information. 

In the numerical results, we show that when the source's transmission capacity $m$ is limited, gossiping can be harmful, i.e., it increases the incorrect information at the network. For the given source's total update capacity, there is an optimal gossip rate that minimizes the average error. When the network size increases, we should also increase both the source's transmission capacity $m$ and the total update rate $\lambda_s$ proportional to $n$. Finally, we show numerically that the adaptive $m$-selection policy indeed outperforms the constant $m$-selection policy even under the high gossip rate regime.

\section{System Model and Problem Formulation}\label{sect:model}
\begin{figure}
	\centering
	\includegraphics[width=0.75\columnwidth]{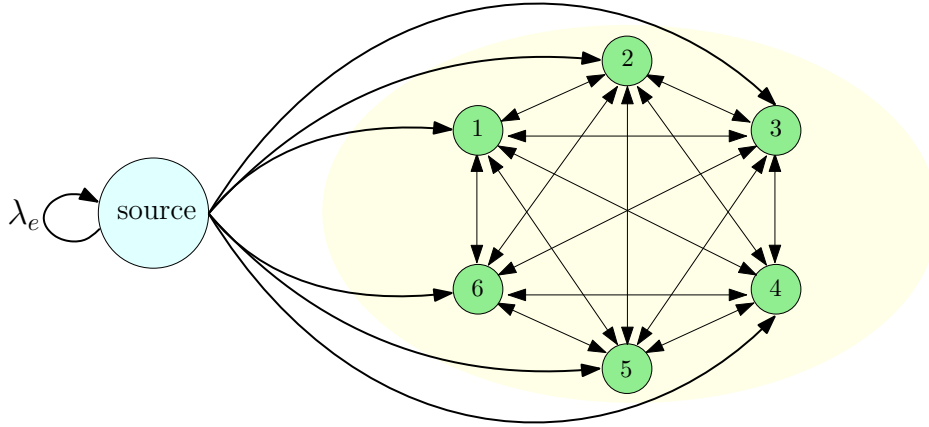}
	\caption{A communication system that consists of a source and fully connected $n$ nodes.}
	\label{Fig:system_model}
\end{figure}

We consider an information updating system consisting of a source and $n$ receiver nodes as shown in Fig.~\ref{Fig:system_model}. The source keeps the most up-to-date information about a state of the world that takes binary values of 0 or 1. The information at the source is updated following a Poisson process with rate $\lambda_e$. We define the time interval between the $j$th and $j+1$th information update at the source as the $j$th update cycle and denote it by $I_j$. We assume that the source is able to send instantaneous signals to the nodes. After receiving these signals, the nodes know that information at the source is updated, but they do not know which information is realized at the source. We denote the information at the source at update cycle $j$ as $x_s(j)$. For a given $x_s(j)$, the information at the source at the $j+1$th update cycle is equal to $x_s(j+1) = x_s(j)$ with probability $1-p$ and $x_s(j+1) =1- x_s(j)$ with probability $p$, i.e.,
\begin{align}
\mathbb{P}(x_s(j\!+\!1)|x_s(j)) \!=\! \begin{cases}
1-p, & \text{if $x_s(j+1) \!=\! x_s(j)$ }, \\
p, & \text{if $x_s(j+1) \!=\! 1\!-\! x_s(j)$},
\end{cases}
\end{align}
for all $j$, where $0<p<1$. 

The source updates each receiver node according to a Poisson process with rate $\frac{\lambda_s}{n}$. In this system, in addition to the update arrivals from the source, each node can share its local information with the other nodes, a process called \textit{gossiping.} Specifically, in this work, we consider a fully connected network where each node is connected to every other node with equal update rates. The total update rate of a node is $\lambda$. Thus, in this network, each node updates other neighbor nodes following a Poisson process with rate $\frac{\lambda}{n-1}$. We denote the information at node $i$ at update cycle $I_j$ as $x_i(j)$. The nodes want to follow the most up-to-date information prevailing at the source as accurately as possible based on the updates that they receive from the source as well as from the neighbor nodes during an update cycle. 

In this paper, we consider an information updating mechanism where at the beginning of each update cycle $I_j$, the source sends its current information to $m$ nodes where $1\leq m\leq n$. Here, we assume that the source knows (or is able to sense/monitor) the information prevailing at the nodes, and thus, it sends updates to the nodes that carry different information compared to the source. During this phase, if the information at the source is updated, then another update cycle starts, and thus the $j$th update cycle can be terminated before sending updates to $m$ nodes. If the source sends updates to $m$ nodes, we enter the gossiping phase in the update cycle $I_j$. During the gossiping phase, the nodes share their local information with each other. When the information at the source is updated, the gossiping phase ends. At the end of the gossiping period, the nodes that did not get an update directly from the source update their information based on the majority of the updates they receive during the gossiping period. If a node does not get any updates from the source or the other nodes, it keeps its local information unchanged. We denote the information at node $i$ at the end of the gossiping period by $x'_i(j)$. In order to measure the performance of the information dissemination process, we define the error metric for node $i$ at the update cycle $j$ as $\Delta_i(j) = x_s(j)-x'_i(j)$.  Then, the average estimation error over all nodes equals $\Delta(j) = \frac{1}{n}\sum_{i=1}^{n} \Delta_{i}(j)$, and the long-term average estimation error over all nodes is given by     
\begin{align}\label{long_term}
   \Delta = \lim_{J\to\infty} \frac{1}{J}\sum_{j=1}^{J} \Delta(j).
\end{align}
In the next section, we provide detailed analyses to characterize the long-term average error $\Delta$.
             
\section{Characterizing the Long-Term Average Error}\label{sect:error_analysis}

In this section, we characterize the long-term average error $\Delta$. Let us consider a generic update cycle $I_j$, and for simplicity of presentation, let us drop the index $j$ from the variables in the rest of the analysis. At the beginning of the update cycle, we denote the number of nodes that have the same information as the source by $N\in\{0,\dots,n\}$. In phase $S$, either the source sends an update to a node after an exponential time with the rate $\lambda_s$ or the information at the source is updated after an exponential time with the rate $\lambda_e$. Thus, the source sends update to a node with probability $\frac{\lambda_s}{\lambda_s+\lambda_e}$ or the information at the source is updated and the next update cycle starts with probability $\frac{\lambda_e}{\lambda_s+\lambda_e}$. Therefore, during a typical update cycle $I$ with $N< n-m$, the source sends $K_s$ updates with the following probability mass function (pmf)  
\begin{align}\label{eqn:pmf_k_s_1}
\mathbb{P}(K_s = k_s| N < n-m ) =\left(\frac{\lambda_s}{\lambda_s+\lambda_e}\right)^{k_s} \frac{\lambda_e}{\lambda_s+\lambda_e},
\end{align}
when $k_s\in\{0,\dots, m-1\}$, and $\mathbb{P}(K_s = m| N < n-m ) =\left(\frac{\lambda_s}{\lambda_s+\lambda_e}\right)^{m}$. Similarly, if $N \geq n-m $, we have
\begin{align}\label{eqn:pmf_k_s_2}
\mathbb{P}(K_s = k_s| N \geq n-m ) =\left(\frac{\lambda_s}{\lambda_s+\lambda_e}\right)^{k_s} \frac{\lambda_e}{\lambda_s+\lambda_e},
\end{align}
when $k_s\in\{0,\dots, n-N-1\}$, and $\mathbb{P}(K_s = n-N| N \geq n-m ) =\left(\frac{\lambda_s}{\lambda_s+\lambda_e}\right)^{n-N}$.

For an update cycle with $N < n-m $, the network enters the gossiping phase with probability $\mathbb{P}(K_s = m| N < n-m ) =\left(\frac{\lambda_s}{\lambda_s+\lambda_e}\right)^{m}$, which decreases with $m$. In other words, choosing a large $m$ decreases the probability of entering the gossiping phase. On the other hand, choosing a small $m$ results in sending updates to a small number of nodes, and thus, in the gossiping phase, incorrect information can be spread. Therefore, there is an optimal $m$ that achieves the smallest average error $\Delta$. 

If the source sends updates to $m$ nodes before the information at the source is updated, then the gossiping phase starts. During the gossiping phase, either each node receives an update from the other nodes after an exponential time with rate $\lambda$ or the information at the source is updated after an exponential time with rate $\lambda_e$. Thus, similar to \cite{bastopcu20e}, during the gossiping phase, node $i$ receives $K_i$ updates with the following pmf:
\begin{align}
\mathbb{P}(K_i = k_i ) =\left(\frac{\lambda}{\lambda+\lambda_e}\right)^{k_i} \frac{\lambda_e}{\lambda_s+\lambda_e}, \ \ k_i = 0,1,\dots.
\end{align}
In other words, $K_i$ has geometric distribution with parameter $\frac{\lambda_e}{\lambda_s+\lambda_e}$, i.e., $K_i\sim {\rm Geo}(\frac{\lambda_e}{\lambda_s+\lambda_e})$. 

At the beginning of the gossiping phase, there are $N+m$ nodes with the same information as the source and $n-N-m$ nodes with incorrect information. For the nodes with $x_i =x_s$, conditioned on the total number of updates $K_i =k_i$ that they received during the gossiping phase, the distribution of the number of updates that are equal to $x_s$ is given by 
\begin{align}\label{eqn_p_r1}
\mathbb{P}(R_i = r| K_i = k_i, x_i = x_s) =
{{k_i}\choose{r}} \left(\frac{N+m-1}{n-1}\right)^{r} \left(\frac{n-N-m}{n-1}\right)^{k_i-r}\!\!\!\!\!\!\!, \ \ \ r = 0,\dots, k_i,
\end{align}
where $R_i$ is a random variable denoting the number of updates that are equal to $x_s$. In other words, for a node $i$ that has $x_i=x_s$, conditioned on $K_i = k_i$, the random variable $R_i$ has a binomial distribution with parameters $(k_i, \frac{N+m-1}{n-1})$, i.e., $R_i\sim {\rm Bin}(k_i, \frac{N+m-1}{n-1})$. Similarly, for the nodes $i$ with $x_i \neq x_s$, we have  \begin{align}\label{eqn_p_r2}
\mathbb{P}(R_i = r| K_i = k_i, x_i \neq x_s) =
{{k_i}\choose{r}} \left(\frac{N+m}{n-1}\right)^{r} \left(\frac{n-N-m-1}{n-1}\right)^{k_i-r}\!\!\!\!\!\!\!, \ \ \ r = 0,\dots, k_i. 
\end{align}  
At the end of the gossiping period, based on the majority of the updates, the nodes $i$ that have $x_s$ as their prior information estimate the information at the source as $x_i' = x_s$ with probability $\mathbb{P}_{T,1}(N)$, which is given by 
\begin{align}\label{eqn_P_T1}
\mathbb{P}_{T,1}(N)  =&  \sum_{k_i = 1}^{\infty} \mathbb{P}(R_i  \geq  \lfloor{\frac{k_i}{2}} \rfloor+1| K_i =  k_i, x_i  =  x_s) \mathbb{P}( K_i  =  k_i)\nonumber\\
&+\frac{1}{2}\sum_{k_i = 1}^{\infty}\mathbb{P}(R_i  =  k_i| K_i  =  2 k_i, x_i  =  x_s) \mathbb{P}( K_i  = 2 k_i)+\mathbb{P}( K_i = 0).
\end{align}
We note that the first summation term in (\ref{eqn_P_T1}) corresponds to the case where a node receives a strictly higher number of $x_s$ during the gossiping period. The second summation term in (\ref{eqn_P_T1}) refers to the case where a node receives equal number of $x_s$ and $1-x_s$. In this case, a node estimates the information as either $x_s$ or $1-x_s$ with equal probabilities. If a node does not get any updates during the gossiping phase, it keeps its current information that is given by the last term in (\ref{eqn_P_T1}). Similarly, for a node $i$ that has prior information $x_i\neq x_s$, we can derive an expression for the probability of updating its information to $x_s$, denoted by $\mathbb{P}_{T,2}(N)$, as 
\begin{align}\label{eqn_P_T2}
\mathbb{P}_{T,2}(N)  =&  \sum_{k_i = 1}^{\infty} \mathbb{P}(R_i  \geq  \lfloor{\frac{k_i}{2}} \rfloor+1| K_i = k_i, x_i \neq x_s) \mathbb{P}( K_i  =  k_i)\cr
&+\frac{1}{2}\sum_{k_i = 1}^{\infty}\mathbb{P}(R_i  =  k_i| K_i  =  2 k_i, x_i \neq x_s) \mathbb{P}( K_i  = 2 k_i).
\end{align}
Note that this expression is identical to that in \eqref{eqn_P_T1}, except that in the summations, we use the probabilities $\mathbb{P}(R_i = r| K_i = k_i, x_i \neq x_s)$ given in (\ref{eqn_p_r2}), and the last term $\mathbb{P}( K_i = 0)$ is excluded.

At the end of an update cycle with gossiping phase, $m$ nodes that obtain information directly from the source will have $x_i' = x_s$.\footnote{In the gossiping phase, these nodes send information to other nodes with rate $\lambda$, but they do not update their information based on the updates received from the other nodes.} There are $N$ nodes that have prior information $x_s$. These nodes will update their information to $x_i' = x_s$ with probability $P_{T,1}(N)$ and to $x_i' = 1-x_s$ with probability $1-P_{T,1}(N)$. Thus, the total number of nodes that update their information to $x_s$, denoted by $N_1'$, has the binomial distribution $N_1'\sim {\rm Bin}(N, P_{T,1}(N))$. On the other hand, there are $n-N-m$ nodes that have prior information $1-x_s$. At the end of the gossiping phase, these nodes will update their information to $x_i' = x_s$ with probability $P_{T,2}(N)$, and to $x_i' = 1-x_s$ with probability $1-P_{T,2}(N)$. Thus, the total number of nodes that change their information to $x_s$, denoted by $N_2'$, obeys the binomial distribution $N_2'\sim {\rm Bin}(n-N-m, P_{T,2}(N))$. Therefore, at the end of the gossiping period, the total number of the nodes that have $x_s$ is equal to $m+N'$, where $N'= N'_1+N'_2$ has the following pmf 
\begin{align}
\mathbb{P}(N' =n') = \sum_{\ell_1 = \ell_{lower}}^{\ell_{upper}}\mathbb{P}(N_1'= \ell_1)\mathbb{P}(N_2'= n'-\ell_1),\ \ \ n' = 0,\dots,n-m,
\end{align}
where $\ell_{lower}=\max\{0,n'+N+m-n\}$ and $\ell_{upper} = \min\{N,n'\}$.

Next, let us define $N''(j)$ to be the number of nodes that have the same information with the source at the end of the update cycle $I_j$, i.e., $x_i'(j) = x_s(j)$. If the update cycle $I_j$ ends before entering the gossiping phase, then either $N(j)< n-m, K_s<m$ or $N(j)\geq n-m$. In these cases, the source sends updates to $k_s$ nodes with probability distributions given in (\ref{eqn:pmf_k_s_1}) and (\ref{eqn:pmf_k_s_2}), respectively. If the source is able to send updates to $m$ nodes, then gossiping phase starts and as a result, $N''(j) = m+n'$ nodes will have $x_s(j)$ with probabilities $\mathbb{P}(K_i = m)\mathbb{P}(N' =n')$, where $n' = 0,\dots,n-m$. Thus, the probability distribution of $N''$ for a given $N$ is equal to 
\begin{align}\label{eqn_P_n''}
\mathbb{P}(N'' \!=\!  n''|N) \!=\! \begin{cases}
\mathbb{P}(K_s = k_s| N < n-m ), & \vspace{-1em}\text{if $n''= k_s+N< m$, }\\ &\text{and $k_s = 0,\dots,m-1$ }, \\
\mathbb{P}(K_s = m| N < n\!-\!m )\mathbb{P}(N' =n'), &  \text{if $m\leq  n''\!=m\!+\!n'<N $}, \\
\mathbb{P}(K_s \!=\! n''\!-\!N| N < n\!-\!m )\vspace{-1em}\\+\mathbb{P}(K_s \!=\! m| N < n\!-\!m )\mathbb{P}(N' =n''\!-\!m),  &\text{if $ m\leq N\leq n''<N\!+\!m $}, \\
\mathbb{P}(K_s = m| N < n-m )\mathbb{P}(N' =n''-m), & \text{if $N+m\leq n''\leq n $}, \\
\mathbb{P}(K_s = n''-N| N \geq n-m ), &\text{if $n-m\leq N\leq n''\leq n $}.
\end{cases}
\end{align}

With the pmf of $N''$ as provided in (\ref{eqn_P_n''}), we can fully characterize the transition probabilities of going from $N$ nodes that have $x_s$ at the beginning of an update cycle to $N''$ nodes that have $x_s$ at the end of that update cycle. Now let us consider a Markov chain over the state space $(x_s,N)$, where by abuse of notation, we label the first $n+1$ states $(0,0),(0,1),\ldots,(0,n)$ by $1,2,\ldots,n+1$, and the last $n+1$ states $(1,0),(1,1),\ldots,(1,n)$ by $n+2,n+3,\ldots,2n+2$. We can then represent the transition probabilities between different states $a,b\in \{1,2,\ldots,2n+2\}$ using a stochastic matrix $\boldsymbol{P}\in \mathbb{R}^{2(n+1)\times2(n+1)}$, where $P_{a,b}$ denotes the probability of moving from state $a$ to state $b$, and is given by
\begin{align}\label{eqn:P_mat}
\!\!\!\!P_{a,b}\! =\! \begin{cases}
(1-p)\mathbb{P}(N'' = b-1| N =a-1 ), & \!\text{if $1\leq a\leq n+1$, $1\leq b\leq n+1$ }, \\
\frac{p}{1-p} P_{a,2n+3-b}, & \!\text{if $1\leq a\leq n+1$, $n+1\leq b\leq 2(n+1)$ }, \\
\frac{p}{1-p} P_{a,2n+3-b}, & \!\text{if $n+1\leq a\leq 2(n+1)$, $1\leq b\leq n+1$ }, \\
(1\!-\!p) \mathbb{P}(N'' \!= \!b\!-\!n\!-\!2| N \!=\!a\!-\!n\!-\!2 ), & \!\text{if $n\!+\!1\!\leq\! a\!\leq\! 2(n\!+\!1)$, $n\!+\!1\!\leq\! b\!\leq\! 2(n\!+\!1)$}.
\end{cases}
\end{align}     
We note that the stochastic matrix $\boldsymbol{P}$ in (\ref{eqn:P_mat}) is irreducible as every state $b$ is accessible from any state $a$ in a finite update cycle duration. Since $P_{a,a}>0$ for all $a$ in (\ref{eqn:P_mat}), the Markov chain induced by $\boldsymbol{P}$ is also aperiodic. Thus, the above Markov chain admits a unique stationary distribution given by the solution of $\boldsymbol{\pi} = \boldsymbol{\pi} \boldsymbol{P}$, where $\boldsymbol{\pi} = [\pi_{0,0},\dots, \pi_{0,n},\pi_{1,0},\dots, \pi_{1,n}]$ is the row vector of steady-state probabilities of being at different states such that $ \sum_{i=0}^{1}\sum_{j=0}^{n}\pi_{ij} = 1$, $\pi_{ij}\geq 0, \forall i,j$. Finally, we can characterize the long-term average error among all the nodes by
\begin{align}
    \Delta = \sum_{j=0}^{n}\sum_{n'' =0}^{n} (\pi_{0j}+\pi_{1j})\mathbb{P}(N'' = n''|N=j) \frac{n-n''}{n}.
\end{align}
In the following section, we proceed to approximate the probabilities $\mathbb{P}_{T,1}(N)$ and $\mathbb{P}_{T,2}(N)$ provided in this section to understand the effect of gossiping better when the gossip rate $\lambda$ is low and high compared to the information change rate at the source $\lambda_e$.

\section{Analysis for High and Low Gossip Rates} \label{sect:P_T_app}
In this section, we develop approximations for $\mathbb{P}_{T,1}(N)$ and $\mathbb{P}_{T,2}(N)$, which are the probabilities of choosing $x_s$ at the end of a gossiping period when the nodes have the same prior information with the source and when they do not, respectively. First, by assuming sufficiently large $n$ and $N$, we can approximate the conditional pmfs for $R_i$ given in (\ref{eqn_p_r1}) and (\ref{eqn_p_r2}) by the binomial distribution $\mathbb{P}(R_i|K_i=k_i)\sim {\rm Bin}(k_i, \frac{N+m}{n})$. Let us denote the corresponding $\mathbb{P}_{T,1}(N)$ and $\mathbb{P}_{T,2}(N)$ obtained by substituting this binomial approximation into \eqref{eqn_P_T1} and \eqref{eqn_P_T2} by $\hat{\mathbb{P}}_{T,1}(N)$ and $\hat{\mathbb{P}}_{T,2}(N)$, respectively. As $\hat{\mathbb{P}}_{T,1}(N) = \hat{\mathbb{P}}_{T,2}(N) +\mathbb{P}(K_i = 0)$, for the rest of this section, we will only approximate $\hat{\mathbb{P}}_{T,2}(N)$, and the probability $\hat{\mathbb{P}}_{T,1}(N)$ can be found accordingly. Next, for sufficiently large values of $k_i$, we can approximate $\mathbb{P}(R_i\geq \frac{k_i}{2}|K_i=k_i)$ as
\begin{align}\label{eqn_app_r_i}
    \mathbb{P}\left(R_i\geq \frac{k_i}{2}|K_i=k_i\right) \approx Q\left(\sqrt{k_i}A(N)\right),
\end{align}
where $A(N) \!=\! \frac{\frac{1}{2}-\frac{N+m}{n}}{\sqrt{\frac{N+m}{n}\left(1\!-\!\frac{N+m}{n}\right)}}$ and $Q(x) \!=\! \frac{1}{\sqrt{2\pi}}\int_{x}^{\infty} e^{-\frac{u^2}{2}}\,du$. We note that (\ref{eqn_app_r_i}) is due to the normal approximation of binomial distribution by using Central Limit Theorem (CLT). In the following lemma, we show that $\hat{\mathbb{P}}_{T,2}(N)$ can be approximated by a summation of $Q$-functions.
\begin{lemma}\label{Lemma_1}
    When $\lambda$ is sufficiently large compared to $\lambda_e$, $\hat{\mathbb{P}}_{T,2}(N)$ can be approximated by
    \begin{align}\label{eqn:Lemma_1}
       \mathbb{P}_{T,app}(N)= \sum_{k_i=1}^{\infty} Q\left(\sqrt{k_i}A(N)\right)\mathbb{P}(K_i = k_i). 
    \end{align}
\end{lemma}

\begin{Proof}
Using the CLT, there exists a sufficiently large $K$ such that the difference between the probabilities $\mathbb{P}(R_i\geq \frac{k_i}{2}|K_i=k_i)$ and $ Q\left(\sqrt{k_i}A(N)\right)$ is smaller than $\epsilon>0$. Then, we have
\begin{align}\nonumber
    \big|\hat{\mathbb{P}}_{T,2}(N)-\mathbb{P}_{T,app}(N)\big| \leq \sum_{k_i=1}^{K} \Big|\mathbb{P}(R_i \! \geq \! \frac{k_i}{2}| K_i\!  =\!  k_i) \! -\! Q\left(\!\sqrt{k_i}A(N)\right)\!\!\Big| \mathbb{P}( K_i\!  =\!  k_i)+ \epsilon \left(\frac{\lambda}{\lambda\!+\!\lambda_e}\right)^{K+1}\!\!,
\end{align}
where $\mathbb{P}( K_i\!  =\!  k_i) =\left(\frac{\lambda}{\lambda+\lambda_e}\right)^{k_i} \frac{\lambda_e}{\lambda+\lambda_e}$, for $k_i = 0,\dots,\infty$. Note that the above expression can be further upper-bounded by
\begin{align}\nonumber
    \big|\hat{\mathbb{P}}_{T,2}(N)\!-\!\mathbb{P}_{T,app}(N)\big| \leq 1-\left(\frac{\lambda}{\lambda\!+\!\lambda_e}\right)^{K+1}\!\!\!\!\!\!+ \epsilon \left(\frac{\lambda}{\lambda\!+\!\lambda_e}\right)^{K+1}\!\!.
\end{align}
Since the term $1-\left(\frac{\lambda}{\lambda\!+\!\lambda_e}\right)^{K+1}$ can be made smaller than $\epsilon$ by choosing $\lambda>\frac{\lambda_e(1-\epsilon)^{1/(K+1)}}{1-(1-\epsilon)^{1/(K+1)}}$, the difference between $\hat{\mathbb{P}}_{T,2}(N)$ and $\mathbb{P}_{T,app}(N)$ can be smaller than $2\epsilon$ for every $\epsilon>0$ by choosing sufficiently large $\lambda$.   
\end{Proof}

In the previous lemma, we showed that $\hat{\mathbb{P}}_{T,2}(N)$ could be approximated by the summation of $Q$-functions when $\lambda$ is sufficiently large. In the following lemma, we show that as $\lambda\to \infty$, the probability $\hat{\mathbb{P}}_{T,2}(N)$ converges to a step function.

\begin{lemma}\label{lemma_2}
As $\lambda\to \infty$, the probability $\hat{\mathbb{P}}_{T,2}(N)$ converges to a step function given by
\begin{align}
    \lim_{\lambda\to\infty} \hat{\mathbb{P}}_{T,2}(N) \approx \begin{cases}
0, & \text{when $\frac{N+m}{n}<\frac{1}{2}$}, \\
\frac{1}{2}, & \text{when $\frac{N+m}{n}= \frac{1}{2}$}, \\
1, & \text{when $\frac{N+m}{n}> \frac{1}{2}$}.
\end{cases}
\end{align}
\end{lemma}
\begin{Proof}
First, we consider the case when $\frac{N+m}{n}<\frac{1}{2}$. In this case,  we note that $Q\left(\sqrt{k_i}A(N)\right)$ is a decreasing function of $k_i$. Thus, for any arbitrary $\epsilon_1>0$, there exists an $L$ such that $Q\left(\sqrt{k_i}A(N)\right)<\epsilon_1,\ \forall k_i>L$. Therefore, we have
 \begin{align}\nonumber
     \hat{\mathbb{P}}_{T,2}(N) < \sum_{k_i =1}^{L} Q\left(\sqrt{k_i}A(N)\right)\mathbb{P}(K_i \!=\! k_i) \!+\!\epsilon_1 \left(\!\frac{\lambda}{\lambda\!+\!\lambda_e}\!\right)^{L+1}\!.   
 \end{align}
 Since $ Q\left(\sqrt{k_i}A(N)\right) <\frac{1}{2}$ for $k_i\geq 1$, as in the proof of Lemma~\ref{Lemma_1}, by choosing sufficiently large $\lambda$, one can show that $ \hat{\mathbb{P}}_{T,2}(N)<2\epsilon_1$. Thus, if $\frac{N+m}{n}<\frac{1}{2}$, we have  $ \lim_{\lambda\to\infty} \hat{\mathbb{P}}_{T,2}(N) = 0 $. 
 
 Next, we consider the case when $\frac{N+m}{n}>\frac{1}{2}$. As $ Q(x) = 1-Q(-x)$ for all $x$, we have 
     \begin{align}\nonumber
       \hat{\mathbb{P}}_{T,2}(N) = \sum_{k_i=1}^{\infty} \left(1-Q\left(-\sqrt{k_i}A(N)\right)\right)\mathbb{P}(K_i = k_i). 
    \end{align}
    Note that $Q\left(-\sqrt{k_i}A(N)\right)$ is a decreasing function of $k_i$. Thus, for any $\epsilon_2>0$, there exists a large $\hat{L}$ such that $Q\left(-\sqrt{k_i}A(N)\right) <\epsilon_2$. Therefore, we can write 
    \begin{align}\nonumber
       \hat{\mathbb{P}}_{T,2}(N)> \frac{\lambda}{\lambda+\lambda_e} -\sum_{k_i =1}^{\hat{L}} Q\left(-\sqrt{k_i}A(N)\right)\mathbb{P}(K_i \!=\! k_i) - \!\epsilon_2 \left(\!\frac{\lambda}{\lambda\!+\!\lambda_e}\!\right)^{\hat{L}+1}. 
    \end{align}
    Now, similar to the first part of the proof, we can show that $\hat{\mathbb{P}}_{T,2}(N)> \frac{\lambda}{\lambda+\lambda_e} -2\epsilon_2$ by selecting a sufficiently large $\lambda$. Thus, when $\frac{N+m}{n}> \frac{1}{2}$, we have $ \lim_{\lambda\to\infty} \hat{\mathbb{P}}_{T,2}(N) = 1 $. Finally, when $\frac{N+m}{n} = \frac{1}{2}$, we note that the $A(N)$ terms in (\ref{eqn:Lemma_1}) become $0$, which implies $\hat{\mathbb{P}}_{T,2}(N)\approx\mathbb{P}_{T,app}(N) =\frac{1}{2}$.     
\end{Proof}

In Lemma~\ref{lemma_2}, we showed that when the gossip rate $\lambda$ is sufficiently large, the nodes start to have access to information from all other nodes. As a result, all the nodes in the network collectively start to behave like a single node where at the end of a gossiping period, the information is updated based on the majority of the information at all nodes. In other words, if the majority of the nodes have the same information with the source, which happens if $\frac{N+m}{2}> \frac{1}{2}$, all the nodes update their information to $x_s$, and thus, they will have the same information with the source at the end of the gossiping period. On the other hand, when the majority of the nodes have the incorrect information $1-x_s$, which happens if $\frac{N+m}{n}<\frac{1}{2}$, all the nodes will have incorrect information at the end of the gossiping period. Therefore, when the information at the source changes frequently (i.e., $\lambda_e$ is large), and the source has limited total update rate capacity (i.e.,  $\lambda_s$ is small), a high gossip rate $\lambda$ can cause incorrect information to disseminate in the network. As a result, gossiping can be harmful in these scenarios. On the other hand, when the source has high transmission rates, at each update cycle, it is enough for the source to send its information to a number of nodes that achieves the majority, i.e., $\frac{N+m}{n}>\frac{1}{2}$. After that, the remaining nodes can obtain the correct information during the gossiping phase. Thus, when the source has enough transmission rate, high gossip rates among the nodes can be utilized by sending the updates to at most half of the network.

Next, we consider the case in which the gossip rate $\lambda$ is relatively low compared to the rate of information change at the source, $\lambda_e$. When the gossip rate is low, the nodes either do not get any updates, in which case they hold on to their prior information, or they mostly get only one update from the other nodes and hence, update their information based on the only received update. In the following lemma, we approximate the probability $\hat{\mathbb{P}}_{T,2}(N)$ when the gossip rate is low. 

\begin{lemma}\label{Lemma_3}
When $\lambda$ is sufficiently small, the probability $\hat{\mathbb{P}}_{T,2}(N)$ can be approximated by
    \begin{align}\label{eqn:Lemma_3}
       \mathbb{P}_{T,app}^{low}(N) = \frac{\lambda}{\lambda + \lambda_e}\frac{N+m}{n}. 
    \end{align} 
\end{lemma}
\begin{Proof}
When $\lambda$ is sufficiently low, the nodes may not receive any updates or receive a single update packet from the other nodes in the gossiping phase. Thus, the nodes that have the incorrect information $1-x_s$ as a prior information obtain $x_s$ with probability $(1-\mathbb{P}(K_i = 0)) \frac{N+m}{n}$, which is equal to 
\begin{align}\nonumber
    \mathbb{P}_{T,app}^{low}(N) = \frac{\lambda}{\lambda + \lambda_e}\frac{N+m}{n}.
\end{align}
Next, we consider the difference between  $\hat{\mathbb{P}}_{T,2}(N)$ and $\mathbb{P}_{T,app}^{low}(N)$, which is given by
\begin{align}\nonumber
    \big|\hat{\mathbb{P}}_{T,2}(N)\!-\! \mathbb{P}_{T,app}^{low}(N)\big| \leq& \sum_{k_i=2}^{\infty} \!\mathbb{P}\!\left(\!\!R_i\!\geq\! \frac{k_i}{2} | K_i \!=\! k_i\!\!\right) \mathbb{P}(\!K_i \!=\! k_i)+ \left(\frac{N+m}{n}\right)\left(\frac{\lambda}{\lambda+\lambda_e}\right)^2.
\end{align}
Since $\mathbb{P}(R_i\geq \frac{k_i}{2} | K_i \!=\! k_i)\leq 1$, we have 
\begin{align}\label{eqn:lemma_3_proof}
    \big|\hat{\mathbb{P}}_{T,2}(N)\!-\! \mathbb{P}_{T,app}^{low}(N)\big| \leq \left(1\!+\!\frac{N\!+\!m}{n}\right)\left(\frac{\lambda}{\lambda+\lambda_e}\right)^2.
\end{align}
Thus, when the gossip rate $\lambda$ is sufficiently low compared to $\lambda_e$, the upper bound on (\ref{eqn:lemma_3_proof}) can be made arbitrarily small, which makes the approximation $\hat{\mathbb{P}}_{T,2}(N)\approx\mathbb{P}_{T,app}^{low}(N)$ tight.  
\end{Proof}

\subsection{Gossip Gain and an Adaptive Policy for Selecting Transmission Capacity}

As a result of gossiping, when $\lambda$ is low, the nodes that have the correct information $x_s$ as prior information keep their information as $x_s$ with probability $\mathbb{P}_{T,app}^{low}(N) + \mathbb{P}(K_i = 0)$, which is given by $\hat{\mathbb{P}}_{T,1}(N)\approx \frac{\lambda}{\lambda + \lambda_e}\frac{N+m}{n} + \frac{\lambda_e}{\lambda + \lambda_e}$. Thus, when $\lambda$ is small, the probability $\hat{\mathbb{P}}_{T,1}(N)$ can be equivalently approximated by 
\begin{align} \label{eqn_p_t_1_approx}
    \hat{\mathbb{P}}_{T,1}(N) \approx 1 -\frac{n-N-m}{n}\frac{\lambda}{\lambda+\lambda_e}.
\end{align}
Therefore, when the gossip rate is low, we have 
\begin{align}\nonumber
    \mathbb{E}[N_1'|N] &\!=\! N\hat{\mathbb{P}}_{T,1}(N) = N- \frac{\lambda}{\lambda+\lambda_e}\frac{N}{n}(n\!-\!N\!-\!m), \cr
    \mathbb{E}[N_2'|N] &\!=\! (n\!-\!N\!\!-\!m)\hat{\mathbb{P}}_{T,2}(N) \!=\! \frac{\lambda}{\lambda\!+\!\lambda_e}\frac{m\!+\!N}{n}(n\!-\!N\!\!-\!m). 
\end{align}
Thus, at the end of the gossiping period, there are $\mathbb{E}[N_1'+N_2'|N]+m = N+m+\frac{\lambda}{\lambda+\lambda_e}\frac{m}{n}(n-N-m)$ nodes that have the same information as the source $x_s$. If we consider the system with no gossiping where only the source can send updates to $m$ nodes, at the end of an update cycle, most $N+m$ nodes have the same information as the source. Thus, compared to the system with no gossiping, the gain (error reduction) obtained as a result of gossiping can be computed as 
\begin{align}\label{gossip_gain}
G(N) = \frac{m}{n^2}(n-N-m) \left(\frac{\lambda}{\lambda+\lambda_e}\right)\left(\frac{\lambda_s}{\lambda_s+\lambda_e}\right)^m,
\end{align}
which is obtained by subtracting $N+m$ from $\mathbb{E}[N_1'+N_2'|N]+m$ and dividing the result by $n$ due to the definition of $\Delta$. Note that the last term $\left(\frac{\lambda_s}{\lambda_s+\lambda_e}\right)^m$ in (\ref{gossip_gain}) is equal to the probability of entering the gossiping phase. 

Let us denote the average error for a system with no gossiping (i.e., $\lambda=0$) by $\Delta_{ng}$. If the gossip rate is low, the overall gain obtained from gossiping, $|\Delta- \Delta_{ng}|$, can be approximated by 
\begin{align}\label{eqn_G_N_total}
     |\Delta- \Delta_{ng}|\approx B(p) \sum_{N=0}^{n-m} (\pi_{0N}+\pi_{1N}) G(N), 
\end{align}
where $B(p)$ is a scaling function in terms of $p$ to represent the effect of gossiping on the steady-state distribution $\boldsymbol{\pi}$. 

When the gossip rate among the nodes is low, the gossip gain $G(N)$ in (\ref{gossip_gain}) depends on the selection of $m$. Therefore, if the source is allowed to dynamically choose its transmission capacity $m$ in terms of $N$, a natural choice is to adaptively select an $m$ which maximizes the gossiping gain by solving $\frac{\partial G(N)}{\partial m}=0$. Solving this equation in terms of $m$ gives us 
\begin{align}\label{m_selection}
m^*(N) = \frac{n-N}{2}\!-\!\frac{1}{\log \rho_s} \!-\! \sqrt{\left(\!\frac{n\!-\!N}{2}\!\right)^2\!+\!\left(\!\frac{1}{\log \rho_s}\!\right)^2},
\end{align}
where $\rho_s = \frac{\lambda_s}{\lambda_s+\lambda_e}$.\footnote{We note that $\frac{\partial G(N)}{\partial m} = 0 $ has two solutions. The other solution is equal to $m^*(N)$ in (\ref{m_selection}) except that the square-root term has positive sign. One can show that this root is always larger than $n-N$ and, thus, can not be a feasible selection for $m$.} In fact, it is easy to see from (\ref{m_selection}) that the optimal solution $m^*(N)$ always lies in the range $0\leq m^*(N)\leq \frac{n-N}{2}$.

When the source has infinite transmission capacity, we have $ \lim_{\lambda_s\to\infty} m^*(N) = \frac{n-N}{2}$, which suggests that the source should send its information to at most half of the nodes that carry incorrect information. In the other extreme case, when the source's transmission capacity is equal to 0, we have $ \lim_{\lambda_s\to0} m^*(N) = 0$, in which case the source should not send its information to any other nodes. In general, for a given $\lambda_s$, $m^*(N)$ in (\ref{m_selection}) is a decreasing function of $N$, which means that when $N$ is small, i.e., when most of the nodes have incorrect information, the source should send updates to a higher number of nodes. As $N$ increases, the source should send updates to a smaller number of nodes as most nodes carry the same information as the source. In the next section, we provide numerical results to shed light on the effects of gossiping on information dissemination.  

\section{Numerical Results} \label{sect:num_res}
 This section has two subsections, where in the first one, we discuss numerical results on the effects of various parameters such as transmission capacity $m$, rate of information change $\lambda_e$, information transmission rate at the source $\lambda_s$, gossip rate $\lambda$, and the number of nodes $n$ on information dissemination in gossip networks, and in the second one, we provide simulation results to corroborate the analytical results in Section~\ref{sect:P_T_app}.


\subsection{Simulation Results for the Effects of Various System Parameters on Information Dissemination}

\begin{figure}
	\centering
	\includegraphics[width=0.75\columnwidth]{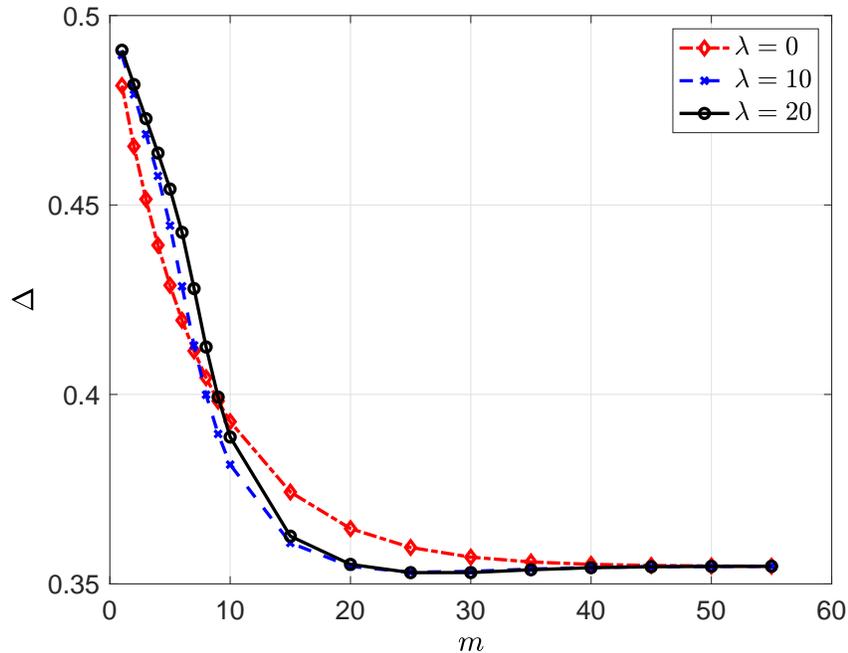}
	\caption{The average error $\Delta$ with respect to $m$ when $\lambda \in \{0,10,20\}$.}
	\label{Fig:sim1}
\end{figure}

In the first numerical result, we take $p=0.4$, $\lambda_e = 1$, $\lambda_s = 10$, and $n =60$. We find the average error $\Delta$ with respect to $m$ when $\lambda = \{0, 10, 20\}.$ Note that $\lambda = 0$ corresponds to the case of no gossiping among the nodes. We see in Fig.~\ref{Fig:sim1} that when $m$ is small, i.e., when the source can send updates to a small number of nodes, the average error $\Delta$ increases with gossip rate $\lambda$. Since $m$ is small and the information change rate $p=0.4$ is high, incorrect information disseminates due to gossiping in the network. As a result, the system with no gossiping ($\lambda= 0$) achieves the lowest average error. When we increase $m$ sufficiently, the nodes start to have access to the same information as the source, and gossiping helps to disseminate the correct information. That is why the systems with gossiping, i.e., $\lambda= 10, 20$, achieve lower average error compared to the system with no gossiping. The lowest average error $\Delta$ is achieved when $m = 25$ for $\lambda = 10, 20$ and $m =55$ for $\lambda=0$. Here, we also note that the average error $\Delta$ is lower when $\lambda=10$ compared to $\lambda=20$, which shows that for a given $m$, there is an optimal gossip rate that achieves the lowest average error. Finally, increasing $m$ further decreases the probability of entering the gossiping phase, and that is why all the curves in Fig.~\ref{Fig:sim1} overlap when $m\geq 40$.                    
\begin{figure}[t]
	\centering
	\includegraphics[width=0.75\columnwidth]{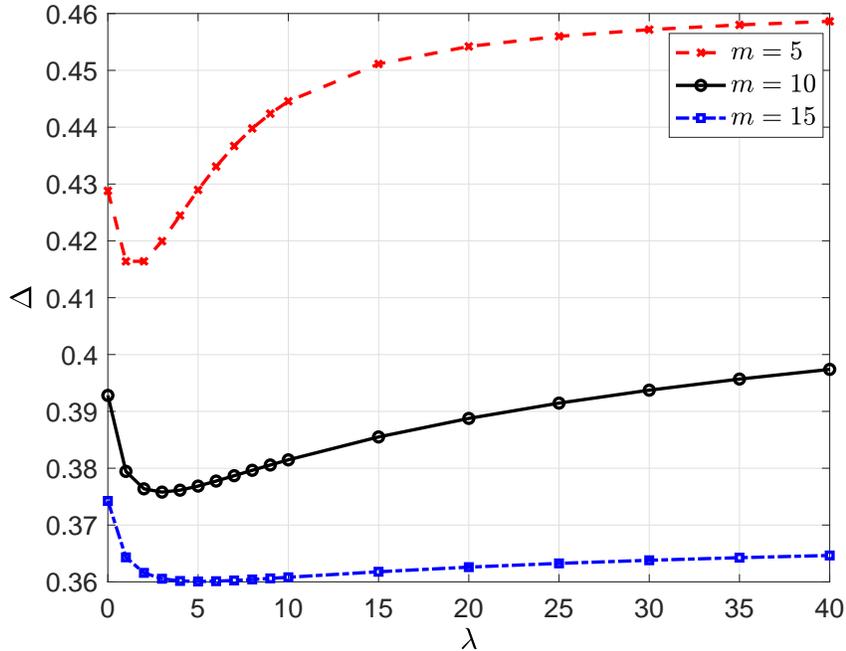}
	\caption{The average error $\Delta$ with respect to the gossip rate $\lambda$ for $m \in \{5,10,15\}$.}
	\label{Fig:sim2}
\end{figure}

In the second numerical result, we consider the same variable selections as in the previous example except that we take $m =\{5,10,15\}$ and change $\lambda$ from 0 to 40. We see in Fig.~\ref{Fig:sim2} that increasing the gossip rate $\lambda$ initially helps to reduce the average error $\Delta$. Then, increasing $\lambda$ further increases $\Delta$ as the incorrect information among the nodes becomes more available. We see in Fig.~\ref{Fig:sim2} that the minimum average error is obtained when $\lambda=1$ for $m=5$, $\lambda = 3$ for $m =10$, and $\lambda =6$ or $\lambda =7$ for $m=15$. We note that as the source sends updates to more nodes, the optimal gossip rate increases.   

\begin{figure}[t]
	\centering
	\includegraphics[width=0.75\columnwidth]{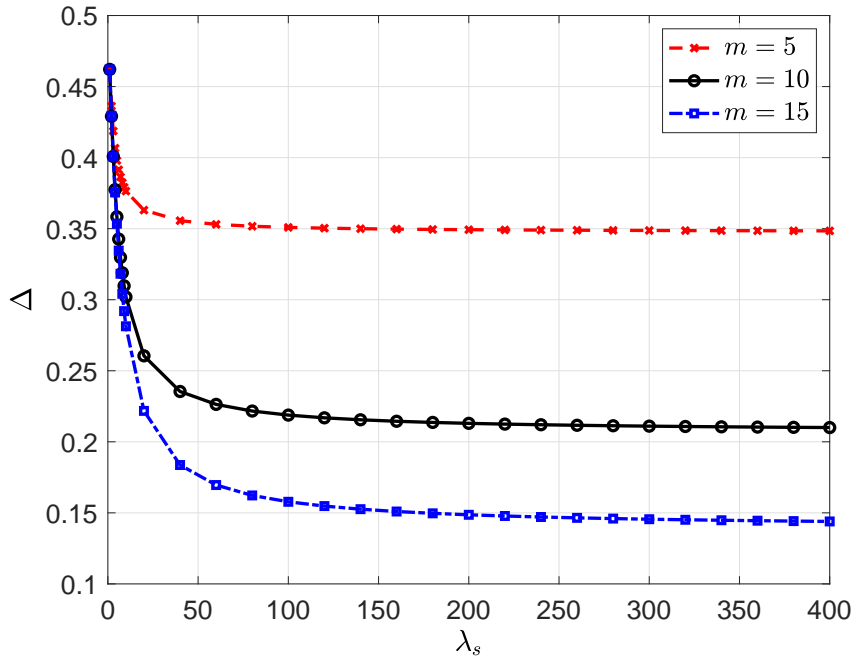}
	\caption{The average error $\Delta$ with respect to the source's update rate $\lambda_s$ for $m = \{5,10,15\}$.}
	\label{Fig:sim3}
\end{figure}

In the third numerical result, we consider $p=0.2$, $\lambda_e = 1$, $\lambda = 5$, and $n =60$. We increase $\lambda_s$ from 1 to 400 for $m =\{5,10,15\}$. We see in Fig.~\ref{Fig:sim3} that increasing $\lambda_s$ initially decreases the average error $\Delta$ faster. However, as $\Delta$ depends also on the other parameters such as $m$ and the gossip rate $\lambda$, increasing $\lambda_s$ further does not improve the average error $\Delta$ and it converges to 0.348 for $m=5$, 0.21 for $m=10$, and 0.144 for $m=15$. 

\begin{figure}[t]
	\begin{center}
    \subfigure[]{\includegraphics[width=0.49\columnwidth]{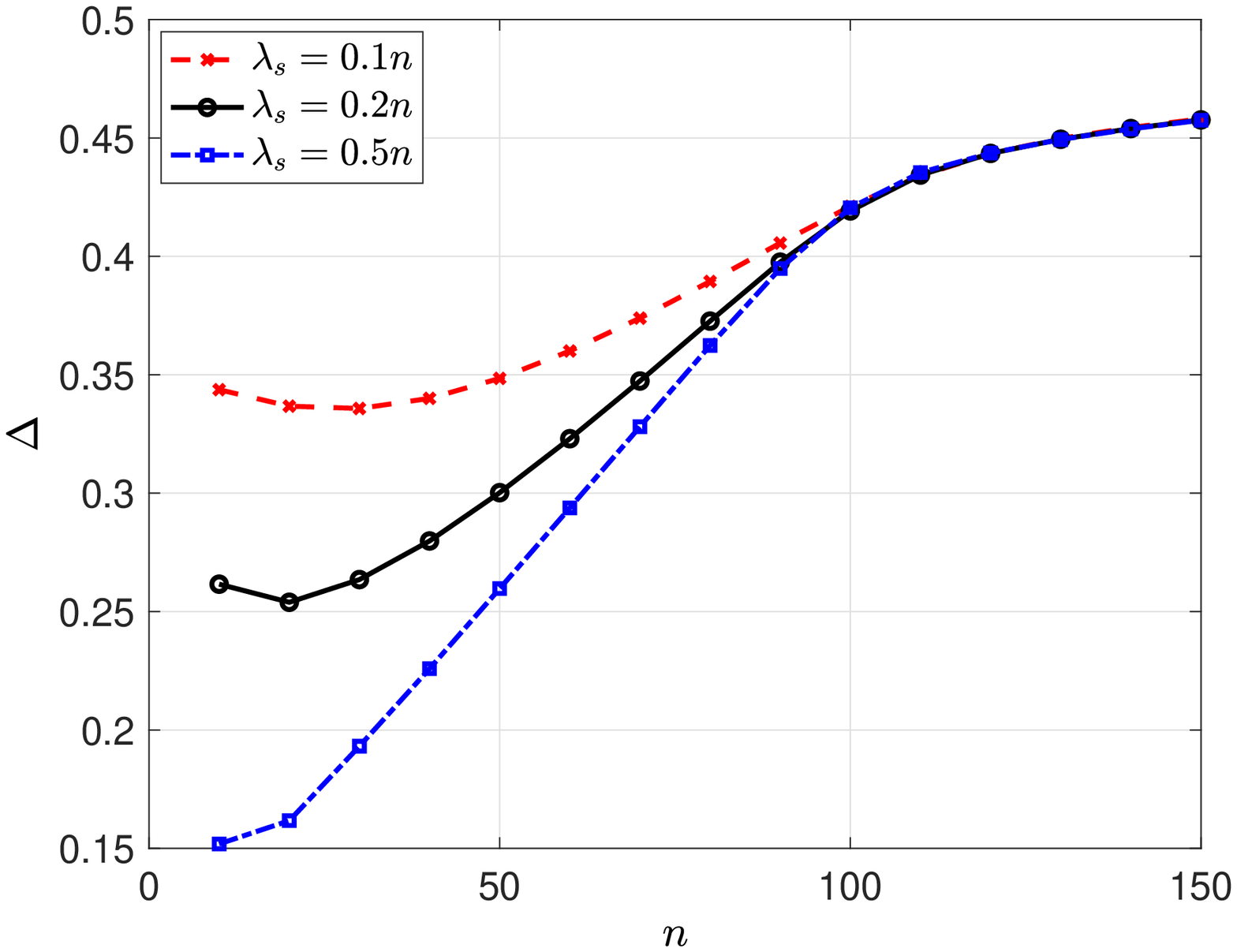}} 
    \subfigure[]{\includegraphics[width=0.49\columnwidth]{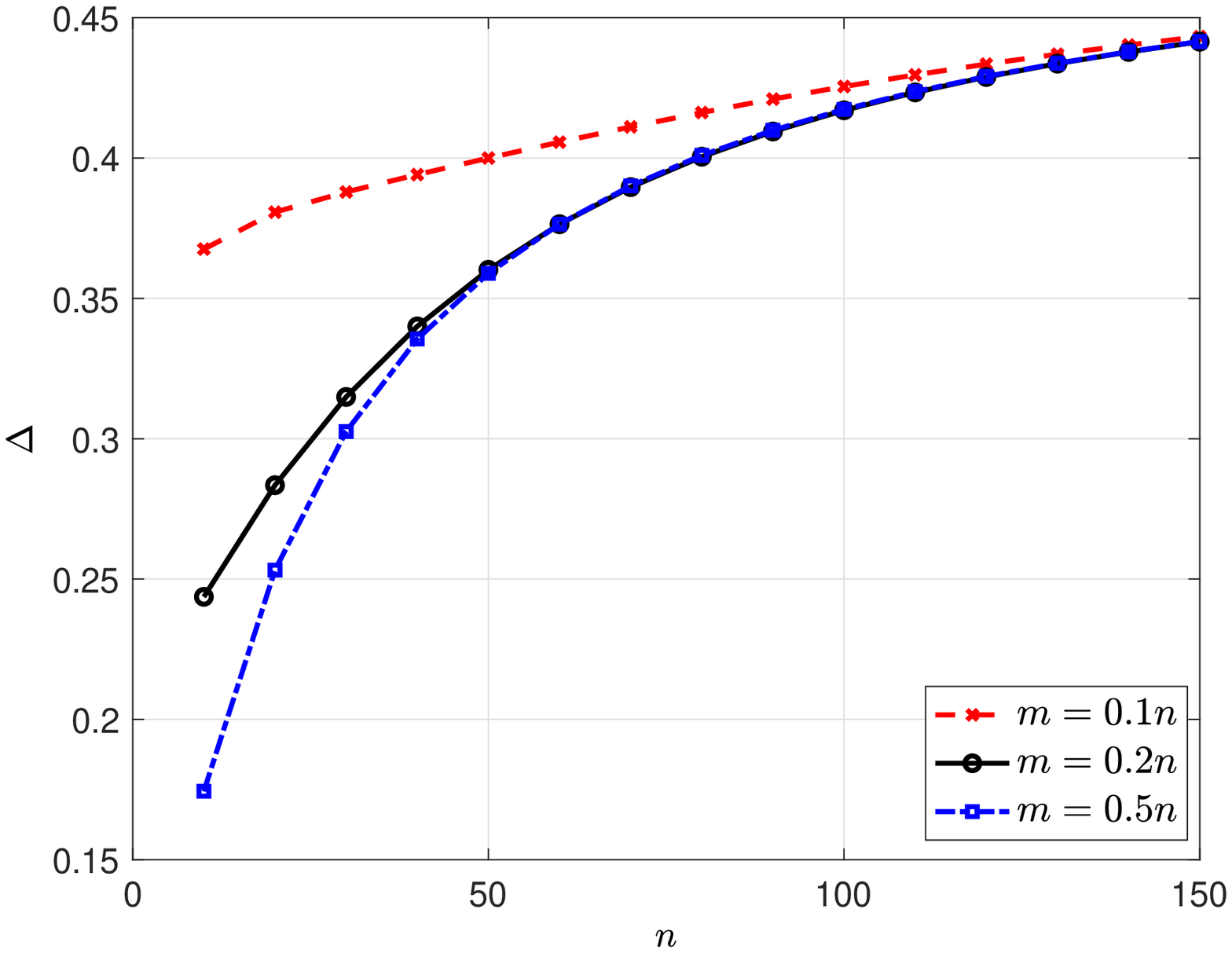}}\vspace{-0.3cm}
    \\ {\subfigure[]{\includegraphics[width=0.49\columnwidth]{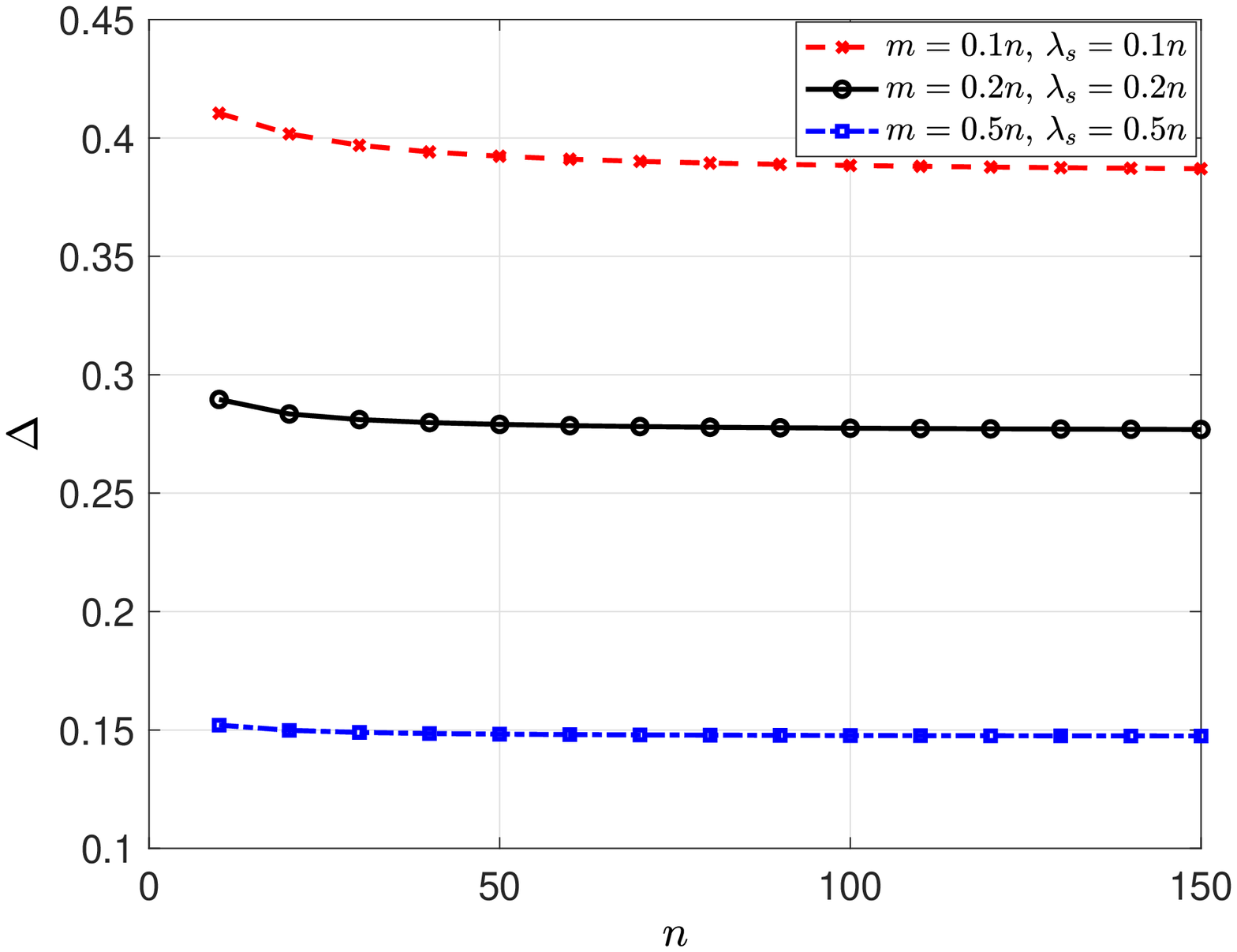}}}
	\end{center}
	\caption{Average error $\Delta$ with respect to $n$ (a) when $\lambda_s \in \{0.1n, 0.2n, 0.5n\}$, (b) when $m \in \{0.1n, 0.2n, 0.5n\}$, and (c) when $m \in \{0.1n, 0.2n, 0.5n\}$, and $\lambda_s \in \{0.1n, 0.2n, 0.5n\}$.}
	\label{Fig:sim4}
\end{figure} 

In the fourth numerical result, we consider the effect of the network size $n$ on the information dissemination. For that, first, we take $p = 0.2$, $\lambda_e = 1$, $\lambda=10$, $m=8$, $n= \{10, 20, \dots, 150\}$ and increase $\lambda_s = \{0.1n, 0.2n, 0.5n\}$ with the network size $n$. In this case, as the network size increases, the source's transmission rate also increases. However, we keep the total number of nodes that the source can send updates to the same, i.e., $m=8$ for all $n$. In Fig.~\ref{Fig:sim4}(a), when $\lambda_s =\{0.1n, 0.2n\}$, we see that the average error $\Delta$ initially decreases with $n$ as $\lambda_s$ is initially a primary limiting factor. Increasing $n$ further increases $\Delta$ as $m$ becomes more important. That is why all these three curves overlap between each other when $\lambda_s$ is sufficiently large. Then, we consider a scenario where we keep $\lambda_s =4$ and only increase $m= \{0.1n, 0.2n, 0.5n \}$. In Fig.~\ref{Fig:sim4}(b), increasing the maximum number of nodes that the source can send updates to in an update cycle alone does not reduce $\Delta$ as $n$ increases. As we increase $n$, $\lambda_s$ becomes the presiding factor and all the curves in  Fig.~\ref{Fig:sim4}(b) overlap. Finally, we increase both the source's transmission rate $\lambda_s$ and capacity $m$ with $n$, i.e., $\lambda_s = \{0.1n, 0.2n, 0.5n\}$ and $m= \{0.1n, 0.2n, 0.5n \}$. As a result, in Fig.~\ref{Fig:sim4}(c), we observe that we can achieve a constant $\Delta$ by increasing $\lambda_s$ and $m$ proportional to $n$.

\subsection{Simulation Results for High and Low Gossiping Rates}

\begin{figure}[t]
	\centering
	\includegraphics[width=0.75\columnwidth]{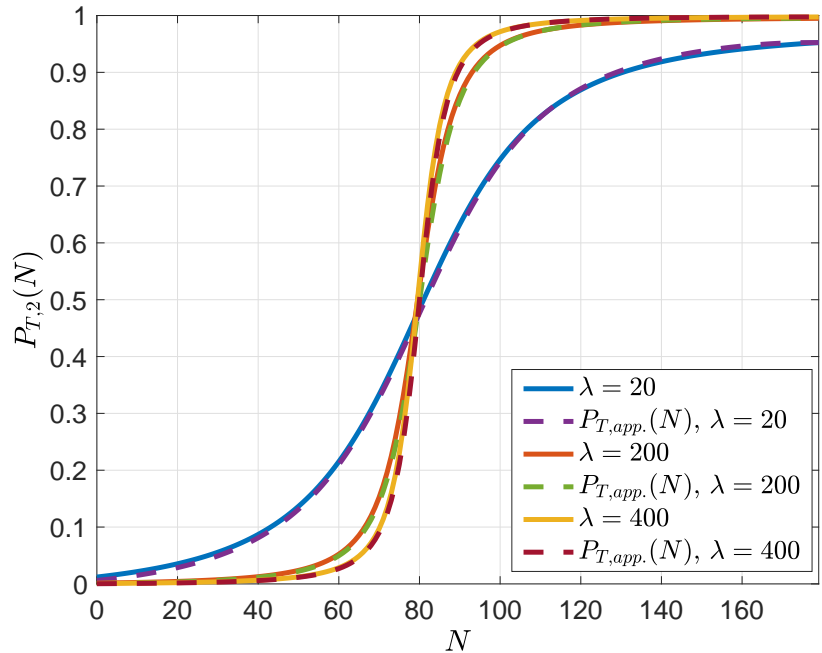}
	\caption{A sample evolution of $\mathbb{P}_{T,2}(N)$ which is approximated by $\mathbb{P}_{T,app}(N)$ in (\ref{eqn:Lemma_1}) when $\lambda$ is high compared to $\lambda_e$.}
	\label{Fig:sim5}
\end{figure}

In this subsection, we provide numerical results for the analysis developed for high and low gossip rates in Section~\ref{sect:P_T_app}. In the first numerical result, we verify the analytical results in Lemmas~\ref{Lemma_1} and \ref{lemma_2}. For this simulation, we numerically evaluate $\mathbb{P}_{T,2}(N)$ when $n=200$, $m=20$, $\lambda_s =2$, $\lambda_e=1$, $p =0.2$ for $\lambda= \{ 20,200,400\}.$ Then, we compare $\mathbb{P}_{T,2}(N)$ with $\mathbb{P}_{T,app}(N)$. In Fig.~\ref{Fig:sim5}, we observe that when $\lambda$ is high compared to $\lambda_e$, $\mathbb{P}_{T,2}(N)$ can be approximated well by $\mathbb{P}_{T,app}(N)$ which is given by the summation of $Q$-functions in (\ref{eqn:Lemma_1}). Furthermore, as we show in Lemma~\ref{lemma_2}, when we increase $\lambda$ from $20$ to $400$, $\mathbb{P}_{T,app}(N)$ and thus $\mathbb{P}_{T,2}(N)$ converge to a step function. Specifically, when $N<\frac{n}{2}-m =80$, we observe that $\mathbb{P}_{T,2}(N)$ converges to $0$ and when $N>\frac{n}{2}-m =80$, $\mathbb{P}_{T,2}(N)$ converges to $1$ while we have $\mathbb{P}_{T,2}(80) =0.5$.      

\begin{figure}[t]
 	\begin{center}
 	\subfigure[]{%
 	\includegraphics[width=0.49\linewidth]{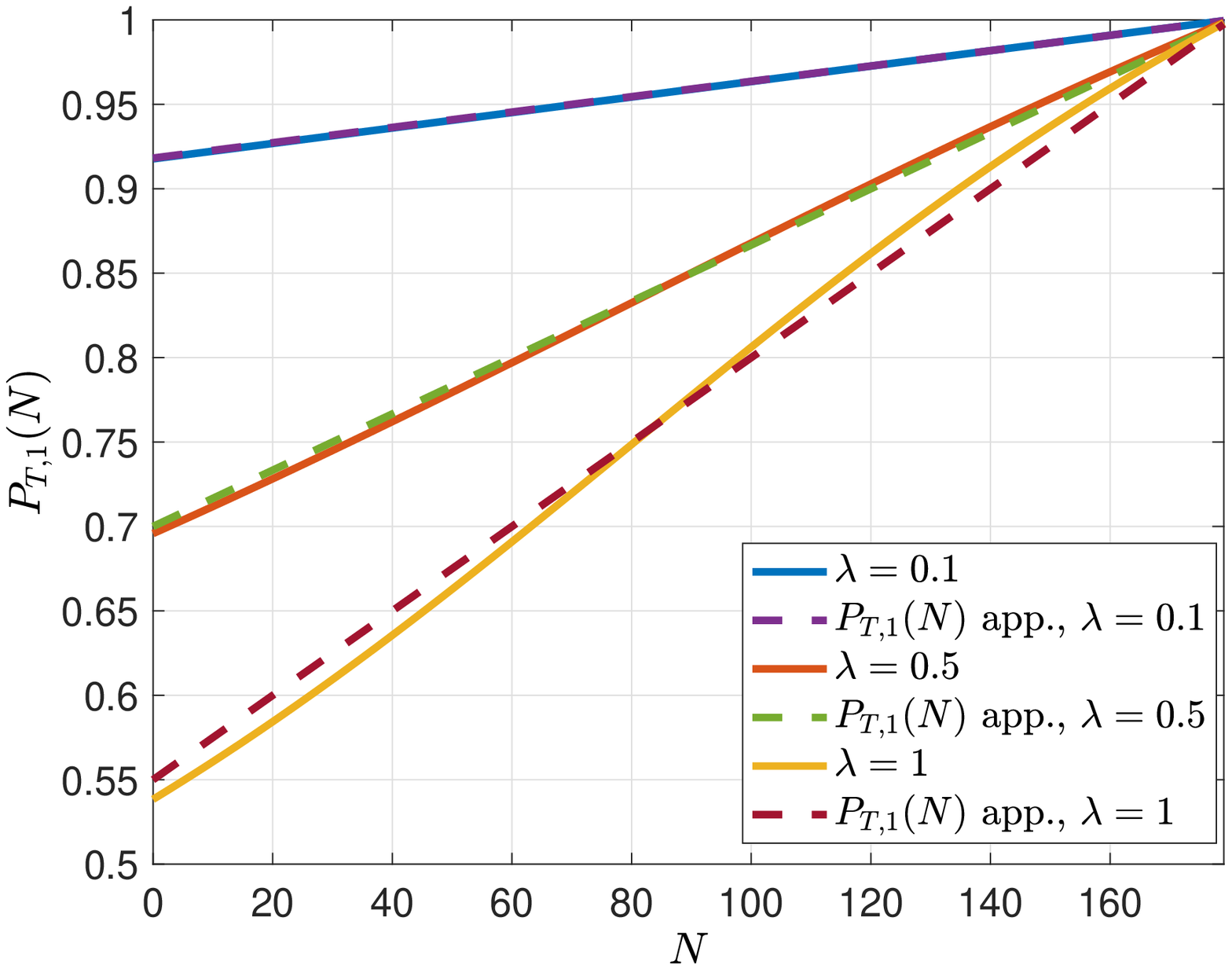}}
 	\subfigure[]{%
 	\includegraphics[width=0.490\linewidth]{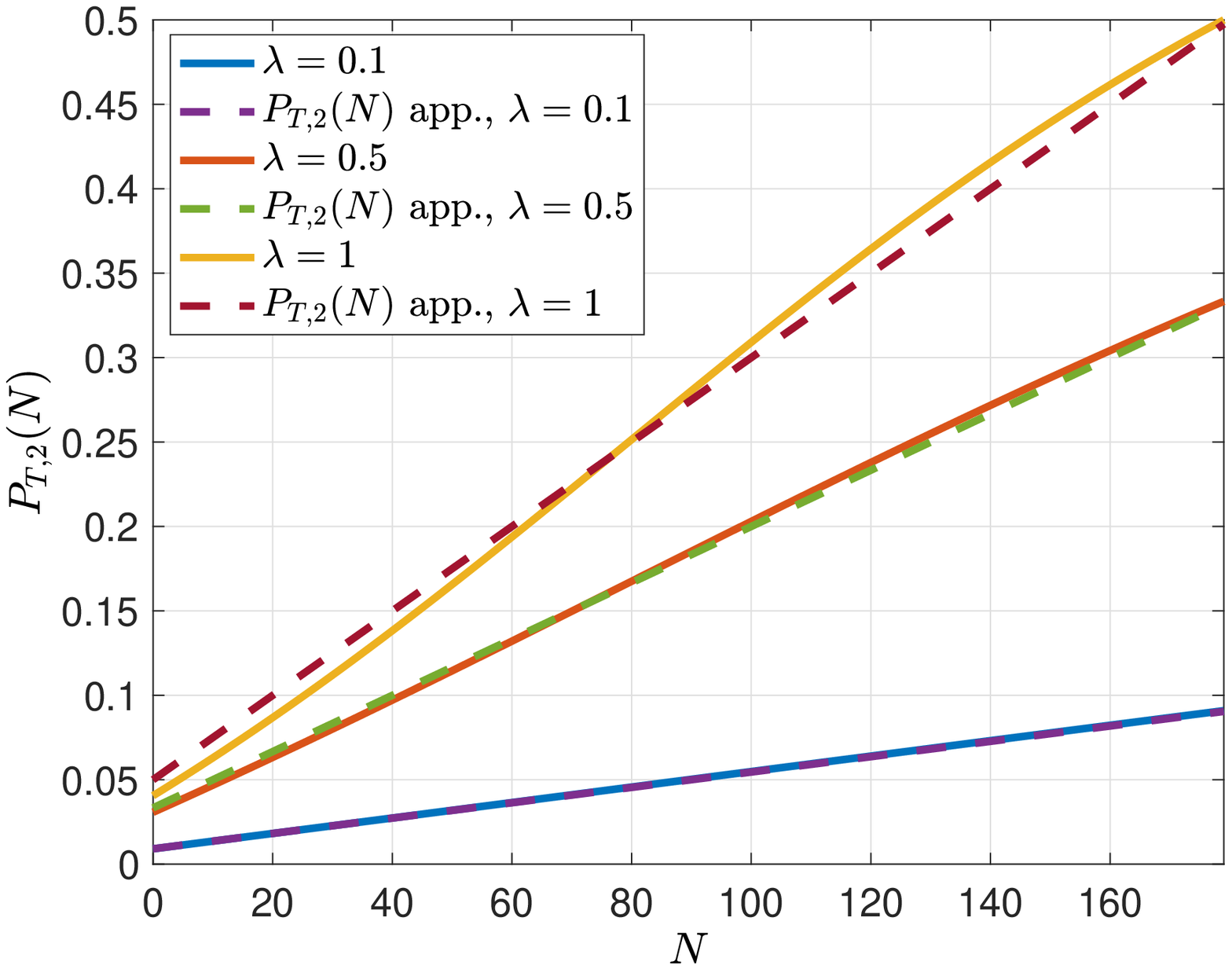}}
 	\end{center}
 	\vspace{-0.4cm}
 	\caption{A sample evolution of $\mathbb{P}_{T,1}(N)$ and $\mathbb{P}_{T,2}(N)$ approximated by (\ref{eqn_p_t_1_approx}) and (\ref{eqn:Lemma_3}) respectively when the gossip rate is low.}
 	\label{Fig:sim6}
 \end{figure}

In the remaining numerical results, we consider the case when the gossip rate $\lambda$ is low compared to $\lambda_e$. In the second simulation result, we evaluate $\mathbb{P}_{T,1}(N)$ and $\mathbb{P}_{T,2}(N)$ with the same parameters except for $\lambda= \{ 0.1, 0.5,1\}.$ We showed in Lemma~\ref{Lemma_3} that when $\lambda$ is low compared to $\lambda_e$, $\mathbb{P}_{T,2}(N)$ can be approximated by $\mathbb{P}_{T,app}^{low}(N)$ in (\ref{eqn:Lemma_3}). We see in Fig.~\ref{Fig:sim6}(b) that when $\lambda = 0.1$ and $\lambda = 0.5$, $\mathbb{P}_{T,2}(N)$ matches closely to $\mathbb{P}_{T,app}^{low}(N)$ in (\ref{eqn:Lemma_3}). When $\lambda = \lambda_e = 1$, $\mathbb{P}_{T,2}(N)$ can still be approximated well by $\mathbb{P}_{T,app}^{low}(N)$, but their differences start to be noticeable. Similarly, for the low gossiping rate, we see in Fig.~\ref{Fig:sim6}(a) that the approximation for $\mathbb{P}_{T,1}(N)$ given in (\ref{eqn_p_t_1_approx}) is close when $\lambda= \{0.1, 0.5\}$. When the gossip rate $\lambda$ is low, during the gossiping phase, the nodes either do not receive any updates, in which case they hold on to their previous beliefs, or only get one update. That is why in Fig.~\ref{Fig:sim6}(b), when $N$ is low, $\mathbb{P}_{T,2}(N)$, which is the probability of having the correct information as a result of gossiping for a node that has incorrect prior information is close to $0$ and then, it increases with $N$.         

\begin{figure}[t]
	\centering
	\includegraphics[width=0.75\columnwidth]{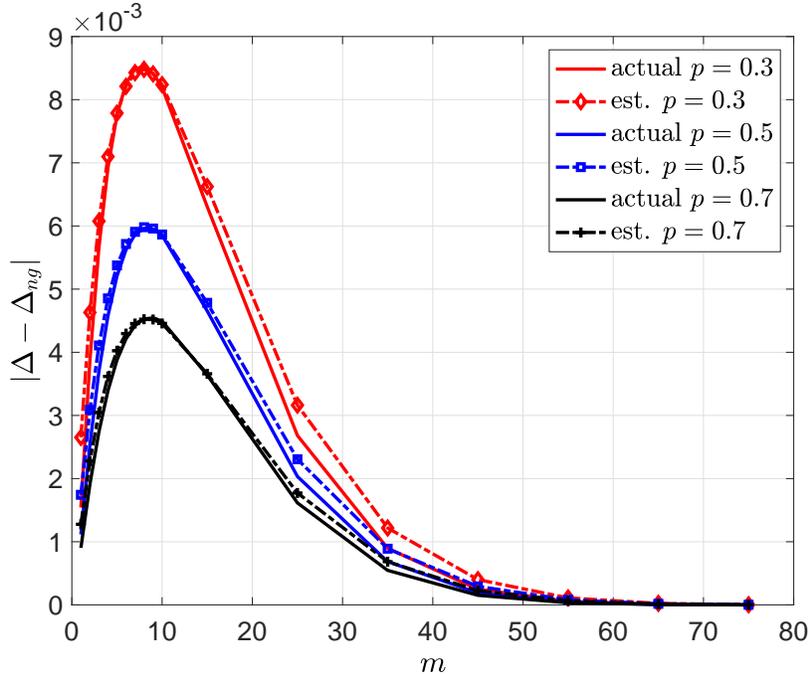}
	\caption{The gossip gain $|\Delta- \Delta_{ng}|$ in (\ref{eqn_G_N_total}) with respect to $m$ for $p = \{0.3, 0.5, 0.7\}$. }
	\label{Fig:sim7}
\end{figure}

In the third simulation result, when the gossip rate is low, we numerically find the gossip gain (\ref{eqn_G_N_total}), which is the difference between the average error with no gossiping $\Delta_{ng}$, and the average error with gossiping $\Delta$. For this example, we take $n=80$, $\lambda=0.4$, $\lambda_s=10$, $\lambda_e = 1$, and $p = \{0.3,0.5,0.7\}$. We plot $|\Delta- \Delta_{ng}|$ with respect to $m$ in Fig.~\ref{Fig:sim7}. We observe in Fig.~\ref{Fig:sim7} that for all values of $p$, the gossip gain initially increases with $m$ as the source sends correct information to a sufficient number of nodes. Then, increasing $m$ further decreases the gossip gain as the probability of entering the gossiping phase decreases in an update cycle. We observe in Fig.~\ref{Fig:sim7} that the optimum gain is obtained when $m = 8$ for all $p$ values. We note that the scaling term $B(p)$ in (\ref{eqn_G_N_total}) is equal to $1.7$, $1.1$, and $0.8$ for $p= 0.2$, $p=0.5$ and $p = 0.7$, respectively. We also note that $G(N)$ in (\ref{gossip_gain}) decreases $N$ in the next update cycle with probability $p$ and increases $N$ with probability $1-p$. Thus, the term $B(p)$ in (\ref{eqn_G_N_total}), which is the amplitude of the gossip gain, decreases with $p$.           

\begin{figure}[t]
	\centering
	\includegraphics[width=0.75\columnwidth]{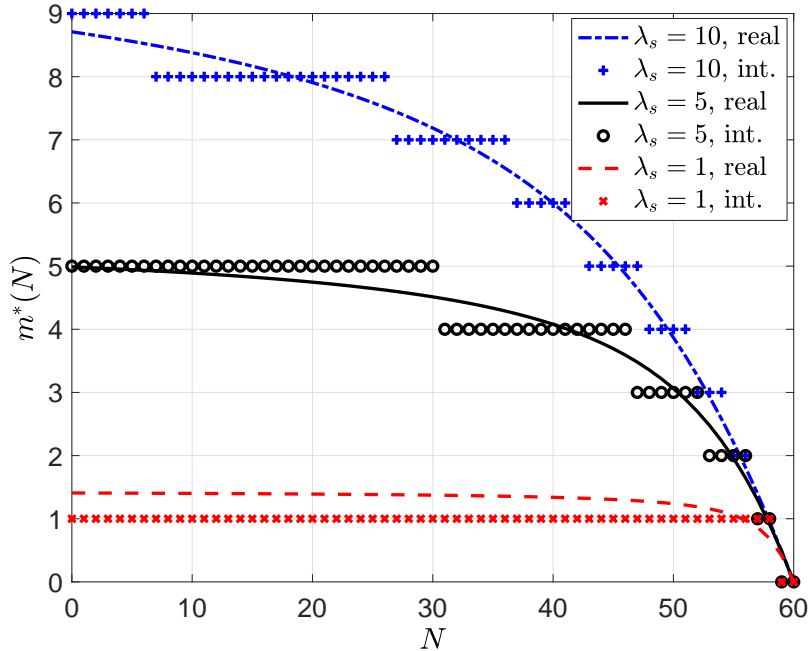}
	\caption{A sample evolution of $m^*(N)$ in (\ref{m_selection}) and its rounding to the nearest integer for different values of $\lambda_s$.}
	\label{Fig:sim8}
\end{figure}

Based on $G(N)$ in (\ref{gossip_gain}), we can find the optimal $m$ that maximizes the gossip gain $G(N)$ for each $N$, which is provided as $m^*(N)$ in (\ref{m_selection}). So far, in this work, we only considered the case where $m$ is kept constant for all update cycles. However, $m^*(N)$ in (\ref{m_selection}) decreases with $N$, which suggests a policy that selects $m$ adaptively depending on $N$. In the next simulation result, we take $n=60$, $p = 0.2$, $\lambda =10$, $\lambda_e = 1$, and $\lambda_s = \{1,5,10\}$. In Fig.~\ref{Fig:sim8}, we plot $m^*(N)$ and their corresponding rounding to the nearest integer. We see in Fig.~\ref{Fig:sim8} that the source sends updates to more nodes as the total update rate of the source $\lambda_s$ increases. 

\begin{figure}[t]
 	\begin{center}
 	\subfigure[]{%
 	\includegraphics[width=0.49\linewidth]{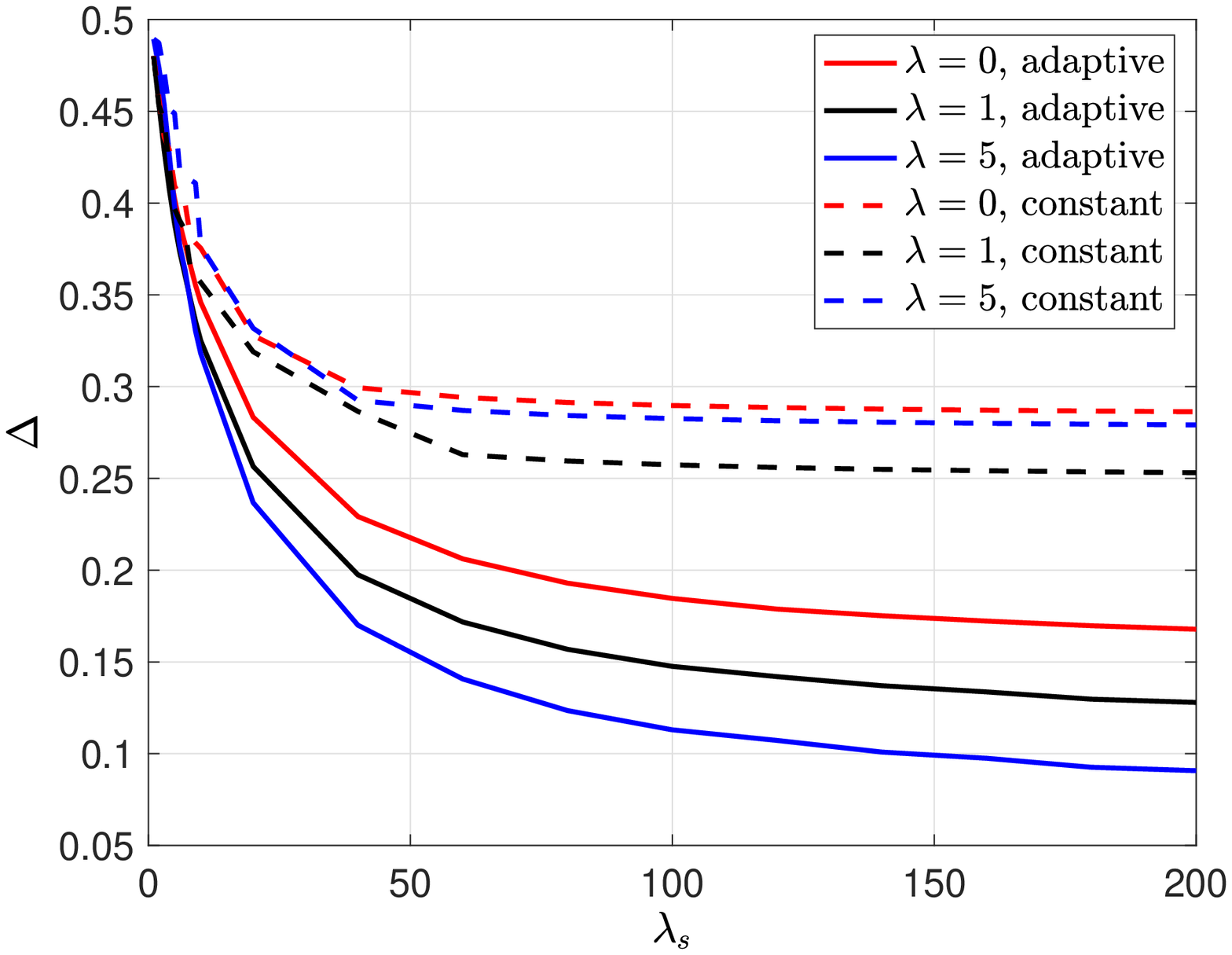}}
 	\subfigure[]{%
 	\includegraphics[width=0.49\linewidth]{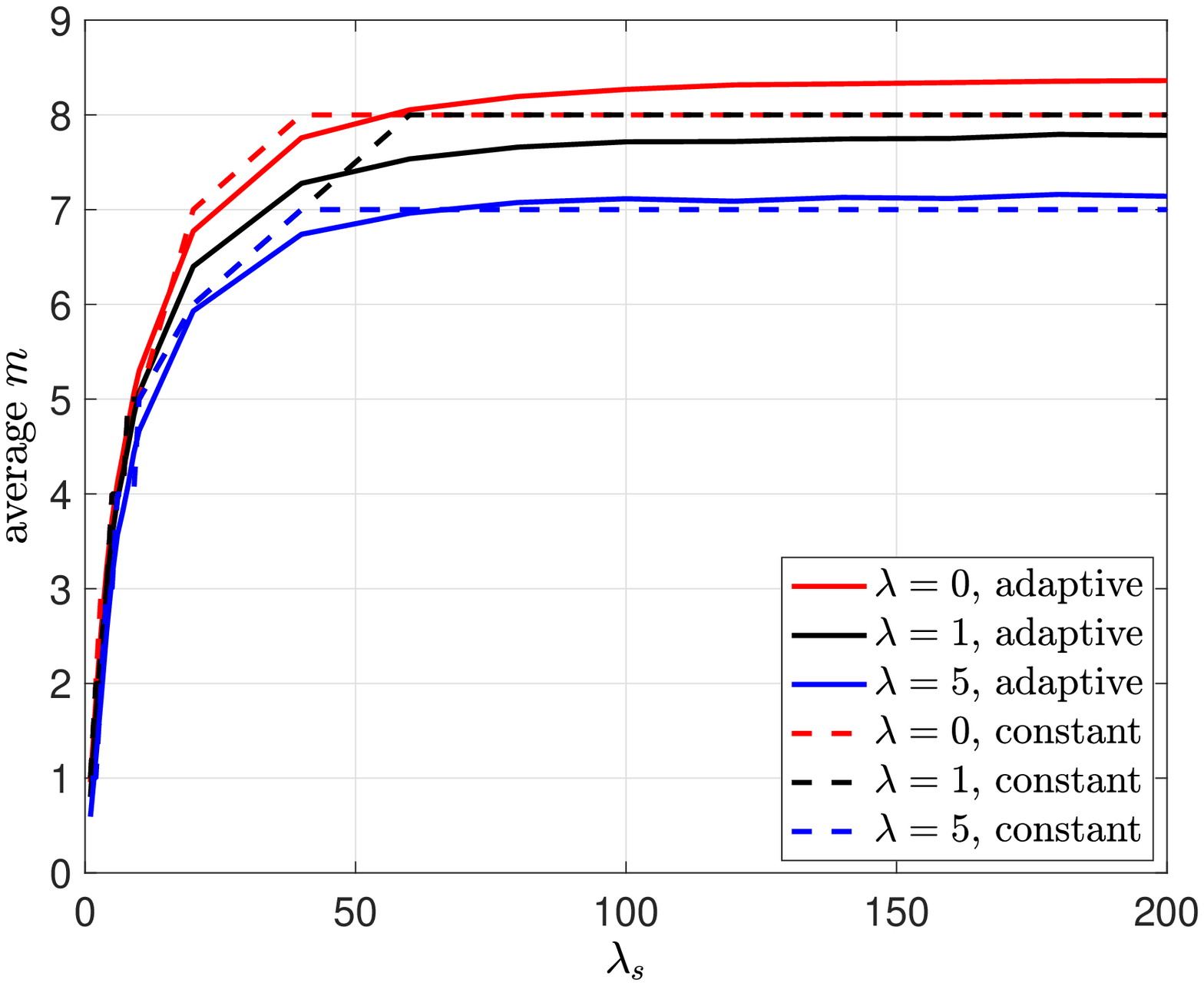}}
 	\end{center}
 	\caption{The comparison between (a) the average error $\Delta$ and (b) the average $m$ for the adaptive $m$ and constant $m$ selection policies.}
 	\label{Fig:sim9}
 \end{figure}

In the last simulation result, we compare the performances of the proposed adaptive policy and the constant policy for selecting $m$. We consider $n =60$, $p = 0.2$, $\lambda =\{0,1,5\}$, $\lambda_e = 1$, and vary $\lambda_s$ from 1 to 200. We first implement the adaptive-$m$ transmission policy by using the nearest integer rounding of $m^*(N)$ in (\ref{m_selection}), which is denoted by $\bar{m}^*(N)$. We then find the stationary distribution $\boldsymbol{\pi}$ and calculate the average $m$ using $\mathbb{E}[\bar{m}^*]=  \sum_{j=0}^{n}(\pi_{0j}+\pi_{1j})\bar{m}^*(N)$, which is depicted in Fig.~\ref{Fig:sim9}(b). In order to make a fair comparison, we take nearest integer rounding of $\mathbb{E}[\bar{m}^*]$, which is shown with the dashed lines in Fig.~\ref{Fig:sim9}(b) and implement the constant $m$ transmission policy. We see in Fig.~\ref{Fig:sim9}(a) that the adaptive $m$ policy (even without gossiping) achieves significantly lower average error $\Delta$ compared to the constant $m$ policy. In Fig.~\ref{Fig:sim9}(a), we also observe that since the gossiping takes place, especially when nodes have the correct information, the average error $\Delta$ decreases with the gossip rate $\lambda$. In the adaptive $m$ selection policy, we see in Fig.~\ref{Fig:sim9}(b) that increasing gossip rate $\lambda$ not only achieves lower $\Delta$, but also decreases the source's transmission capacity $\mathbb{E}[\bar{m}^*]$. Even though we find this policy for low gossip rates ($\lambda<\lambda_e$), we observe that it is an effective transmission policy even for the higher values of $\lambda$ and can achieve lower $\Delta$ compared to the constant $m$ policy.

\section{Conclusion and Future Directions}\label{sect:rest_upt}

In this work, we considered information dissemination over gossip networks consisting of a source that keeps the most up-to-date information about a binary state of the world and $n$ nodes whose goal is to follow the binary state of the world as accurately as possible. In this system, we considered the setting where the source has a limited transmission capacity $m$ and total transmission rate $\lambda_s$. After the source's updates, in order to disseminate the information further, the nodes share their local information with each other. At the end of the gossiping period, the nodes estimate the information at the source based on the majority of the updates that they received. At the end of an update cycle, an estimation error occurs if a node has different information than the source's information.  

We first characterized the equations necessary to obtain the average error $\Delta$ over all nodes. Then, we provided analytical results for the high and low gossip rates. As the information becomes available among the nodes in the high gossip rates, all the nodes behave like a single node. When the gossip rate is low, the nodes either do not get any updates, in which case they keep their prior information or receive a single update. In the low gossip case, we analyzed the gossip gain, which is the error reduction compared to the system with no gossiping, and found $m^*(N)$ that maximizes the gain. That suggests an adaptive $m$ selection policy using $m^*(N)$ where the source sends updates to more nodes if most of them have incorrect prior information. In numerical results, we observed that gossiping could be harmful when the source's transmission capacity $m$ is limited. Moreover, for a given $m$, we numerically determined the optimal gossip rate that minimizes the average error $\Delta$. When the network size $n$ increases, in order to keep $\Delta$ the same, both the source's transmission capacity $m$ and transmission rate $\lambda_s$ need to be increased proportionally to $n$. Finally, we observed that the adaptive $m$ selection policy outperforms the constant $m$ selection policy by achieving significantly lower average error $\Delta$.         

As a future direction, one can consider the problem where the information at the source can take $k>2$ different values based on a known pmf. In this work, we only considered fully connected networks. Extending this work to arbitrarily connected networks could be an interesting direction. To further improve the average error $\Delta$, the idea of clustering networks can be investigated. Finally, one can extend our work to the setting in which the source does not know the prior information of the nodes and thus, has to select $m$ nodes at random. 

\bibliographystyle{unsrt}
\bibliography{IEEEabrv, myLibrary_bastopcu}
\end{document}